\documentclass[aps,reprint,nofootinbib]{revtex4-1}
\usepackage{graphicx,amsfonts,amsmath,amssymb,dcolumn,bm,url}

\begin{document}
\title{Classical Mechanics from Energy Conservation or: Why not Momentum?}
\date{\today}
\author{C. Baumgarten}
\affiliation{Switzerland}
\email{christian-baumgarten@gmx.net}

\def\begeq{\begin{equation}}
\def\endeq{\end{equation}}
\def\bquo{\begin{quotation}}
\def\equo{\end{quotation}}
\def\begary{\begeq\begin{array}}
\def\endary{\end{array}\endeq}
\newcommand{\myarray}[1]{\begin{equation}\begin{split}#1\end{split}\end{equation}}
\def\bmtx{\left(\begin{array}}
\def\emtx{\end{array}\right)}
\def\d{\partial}
\def\h{\eta}
\def\w{\omega}
\def\W{\Omega}
\def\s{\sigma}
\def\eps{\varepsilon}
\def\e{\epsilon}
\def\a{\alpha}
\def\b{\beta}
\def\g{\gamma}
\def\y{\gamma}
\def\d{\partial}
\def\S{{\Sigma}}

\def\leftD#1{\overset{\leftarrow}{#1}}
\def\rightD#1{\overset{\rightarrow}{#1}}

\begin{abstract}
  It is demonstrated that energy conservation allows for a straight derivation
  of Newtonian mechanics without an apriori definition of the concept of work.
  Furthermore it is shown that energy must be depicted as a function of
  position and momentum in order to obtain the correct relativistic equations.
  Accordingly it is argued that not only quantum theory but also special
  relativity is intrinsically a Hamiltonian theory which requires a description
  of the dynamics using coordinate and momentum instead of velocity.

  Furthermore it is argued that the usual historical order of the ``formulations''
  of mechanics, from Newtonian via Lagrangian to Hamiltonian mechanics, is illogical and
  misleading. We suggest that it should be reversed.
\end{abstract}
\maketitle

\section{Introduction}

\bquo
It is as if we developed a kind of schizophrenic scientific personality, using for our teaching,
approaches, concepts, terminologies, examples that our research activities at the same time prove
to be unnecessary, irrelevant or plainly wrong.\\
\mbox{}\hfill{\it -- Jean-Marc L\'evy-Leblond}~\cite{Leblond1976}
\equo

The question whether and how it would be possibly to introduce
mechanics from the concept of energy and energy conservation,
received renewed interest in the recent
years~\cite{Hanc2004,Solbes2009,Brewe2011,Scherr2012a,Scherr2012b,LeGresley2019,Lacy2022}.
It has been suggested in this context to ignore traditional wisdom
and to relate energy conservation to a ``state function'' of
position and velocity~\cite{Carlson2016} and to ignore the well-known
methods of Hamiltonian mechanics. This attempt failed~\cite{Baumgarten2024},
because energy is (in general) a function of position and momentum.
As we shall demonstrate, if energy is treated as a function of position and
velocity, this choice is restricted to the special case of classical
``non-relativistic'' mechanics~\footnote{It has been pointed out by
  L{\'e}vy-Leblond that the notion of ``non-relativistic'' mechanics is
  a misnomer since classical mechanics is subject to Galilean relativity and hence is,
  in principle, ``relativistic'' as well~\cite{Leblond1976}. But to avoid misunderstandings,
  we call a theory relativistic only when it is Lorentz-covariant.}.
We believe that our considerations prove that not only quantum but
also ``relativistic'' (not in the Galilean but in the Einsteinian sense)
mechanics has it's simplest and most natural expression in the
Hamiltonian formalism.

\section{A Quantitative Concept of Work}
\label{sec_work}

There is the tale of Galileo climbing up the tower of Pisa
for an experimental test of free fall~\cite{GUNTHER1935}.
The story tells that the experiments provided evidence that
free fall is independent of the mass (and other properties)
of the falling objects. Whether or not the experiment took
place at the tower of Pisa or not, Galileo's clearly explained
in his {\it dialogues}~\cite{GalileoDialogue} that the
acceleration of free falling bodies in vacuum does not
depend on the mass of the body and he also explained
that a body elastically reflected at ground would return to
the same height. Though Galileo did not introduce the concept
of energy conservation in this context, this would have
been possible.

If Galileo would have had the appropriate knowledge of (vector-)
calculus (which was later invented by Newton and Leibniz),
he could have derived Newton's laws from the simple idea
that the sum of the work stored in position $V(x)$ and the
work stored in motion must be constant, provided that 
friction can be neglected.
Since motion is described by the velocity $v={dx\over dt}$,
the corresponding work stored in motion $T(v)$ should depend
exclusively on the velocity -- or on some other yet undefined
``quantity of motion'', which would vanish if a body is at rest
and increase when the velocity increases.

It has been suggested to ban the concept of work from
physics~\cite{Hilborn2000}, which lead to a short
but not very fruitful controversy~\cite{Mendelson2003,Hilborn2003}.
We claim to the contrary that physical work, introduced in the right
way, provides a simple and clear starting point for the introduction to
mechanics. As we shall demonstrate, there is no need to have
any apriori definition of work. In his famous essay on the principle
of energy conservation, Max Planck wrote that ``the concept of energy only
gains its meaning for physics through the principle which contains it''~\cite{Planck1908,Planck1908en},
which basically means that the concept of energy is inherently defined by the
principle of energy conservation itself. We shall elaborate on
Planck's idea that energy is defined by the principle which
contains it. If Planck is correct than it suffices to acknowledge
the following three facts:
a) It is work to lift a body of a certain mass to some height $x$.
b) It is work to accelerate a body from rest to a certain velocity $v$.
c) The pendulum~\footnote{
It might still be a matter of discussion which mechanical
example is best suited for education, whether it is more
effective to use a pendulum or a spring-mass-system. But
that is not our concern here.
} demonstrates that the work of motion and the
work of position are, in the process of oscillation, transformed
into each other, but that the sum of both types of work is conserved.
To set a pendulum in motion, we can either invest into positional
or in motional work.

\subsection{Energy as a Function of Position and Velocity}
\label{sec_xv}

Hence we suggest to start from the observation that stored
work appears in (at least) two forms, namely in the form of
positional work $V(x)$ and in the form of motional work $T(v)$.
These two forms can be transformed into each other but the
total work, the sum of the two contributions, is (approximately)
conserved. We may then express this condition of the conservation
of work (or energy $E$, respectively) by
\myarray{
  E(x,v)&=T(v)+V(x)\\
  {dE\over dt}&={\d E\over \d v}\,{dv\over dt}+{\d E\over \d x}\,{dx\over dt}=0\\
 &={d T\over d v}\,{dv\over dt}+{d V\over d x}\,v=0\,.
}
Division of the last line by the velocity $v$ and acceleration $a={dv\over dt}$ gives:
\begeq
    {d T\over d v}/v=-{d V\over d x}/a\,.
    \label{eq_1}
\endeq
The left side of Eq.~\ref{eq_1} is exclusively a function of the
velocity (and possibly of properties of the moving object), but
the velocity does not appear on the right side. Hence, if the
equation holds for all velocities, then both sides must be constant.
Let's represent this constant by $m$:
\myarray{
  m&={1\over v}\,{d T\over d v}\\
  m&=-{{d V\over d x}\over {dv\over dt}}\\
}
The first equation can be directly integrated:
\myarray{
  {d T\over d v}&=m\,v\\
  T&=\frac{1}{2}\,m\,v^2+C\\
}
where the integration constant $C$ can be set to zero, since
the energy of motion must be zero if the body is at rest.
The (classical) work stored in motion is then uniquely determined.

The second equation gives with the acceleration $a={dv\over dt}$:
\begeq
  m\,a=-{d V\over d x}\,.
\endeq
If these quantities (left and right) are identified as forces,
then there is a force $F=-{d V\over d x}$ due to the position
dependence of energy which equals the force that is required
to accelerate:
\begeq
F=m\,a\,.
\endeq
From this one derives the definition of the positional work $V(x)$:
\begeq
V=\int\,dV=\int\,F\,dx\,,
\endeq
where the sign convention might still be a matter of discussion.
Our Ansatz demonstrates that there is no need to provide an
apriori definition of work. It suffices to show that the sum
of positional and motional work is a conserved quantity. The
definition of both, kinetic and potential energy, follows automatically.

Hence (mechanical) energy conservation directly and inevitably leads
to Newtonian mechanics with $F=m\,a$ and $T={m\over 2}\,v^2$, whenever
the energy of motion can be treated as a function of velocity.
This means that Newton's second law can be obtained~\footnote{
  It has been questioned that $F=m\,a$ is actually due to Newton,
  but that is not our concern here~\cite{LopesCoelho2025}.
} from the principle of energy conservation.
Unfortunately this also implies that energy conservation alone
apparently does not suffice to obtain the correct (``relativistic'')
equations of motion.

\subsection{Energy as a Function of Position and Momentum}
\label{sec_xp}

In the previous case (Sec.~\ref{sec_xv}) we simply presumed that
position, velocity and acceleration can be treated as independent
variables. This seemingly innocent assumption provided the
possibility to derive a definite expression for the kinetic
energy $T={m\over 2}\,v^2$, independent of the functional
form of the positional energy (i.e. of the potential).

But this can not possibly be the most general of all cases
as we shall demonstrate next. In what follows we shall use 
the same procedure, but merely assume that the work stored
in motion must depend on some (yet undefined) {\it quantity
  of motion} $p$. This quantity of motion must of course
vanish for zero velocity. Thus we have $p(v=0)=0$ and also
$T(p=0)=0$. Furthermore we assume that the quantity of motion
$p$ increases (decreases) whenever the velocity increases
(decreases), i.e. $p(v)$ must be a monotonic function of $v$.

Then one may proceed as before: The energy $E$ is the
sum of two contributions, namely the work stored in position
$V(x)$ and the work stored in motion $T(p)$:
\myarray{
  E(x,p)&=T(p)+V(x)\\
  {dE\over dt}&={\d E\over \d p}\,{dp\over dt}+{\d E\over \d x}\,{dx\over dt}=0\\
  {dE\over dt}&={d T\over d p}\,{dp\over dt}+{\d V\over \d x}\,v=0\\
}
again with $v={dx\over dt}$.
Yet again we devide by $v$ and $\dot p={dp\over dt}$ and obtain
\begeq
    {d T\over d p}/v=-{\d V\over \d x}/\dot p\,.
    \label{eq_2}
\endeq
The left side depends (again) on $v$ and $p(v)$ alone. Hence it
must be equal to some (non-zero) constant.
Since the scale and unit of $p$ are not yet fixed, the constant
can be chosen freely. We choose it to be equal to one and obtain
\begeq
  {dp\over dt}=-{\d V\over \d x}\,.
\endeq
and a general definition of velocity
\begeq
v={d T\over d p}\,,
\label{eq_vdef}
\endeq
which is essentially the first of Hamilton's equations of motion.
But in doing so we do not anymore obtain a definite expression for $T(v)$
or $T(p)$, which implies that this approach covers a wider range of
possibilities and is not limited to Newtonian mechanics. The mere
fact that a more general case is possible already implies that energy
must be regarded as a function of coordinate and momentum $p(v)$
{\it unless} additional information would be available. 
But it also means that we don't have the complete solution, since
the last equation cannot be directly integrated as long as we don't
know how $p$ and $v$ are related in general.

It is clear that $T(p)$ (or $T(v)$, respectively) must be an even
function of it's argument, namely because the
sign of the ``momentum'' $p$ may not alter $T(p)$. Hence in
a Taylor series the lowest order term of relevance for the
equations of motion must be the quadratic term.
The approximation for low velocities therefore starts with the
second order term:
\begeq
T(p)=\frac{a_2}{2!}\,p^2+\frac{a_4}{4!}\,p^4+\dots
\endeq
Combined with Eq.~\ref{eq_vdef} this yields:
\begeq
v={d T\over d p}=a_2\,p+\frac{a_4}{6}\,p^3+\dots
\endeq
The first non-trivial term, that can be used in good approximation, is 
\begeq
v\approx a_2\,p
\endeq
so that, as before, only the determination of the constant
$a_2=m^{-1}$ is left as the remaining step to arrive at
Newton's ``laws''. Hence it is possible to ``derive'' Newtonian
mechanics from energy conservation, even if one uses momentum
and not velocity as one of the two independent and primary dynamical
variables.

\section{``Relativistic'' Mechanics}
\label{sec_str}

\bquo
It is clear that "the (special) theory of relativity" is badly misnamed
from at least two points of view. First, it is {\bf a}, and not {\bf the}
theory of relativity. A perfectly consistent theory of relativity
underlies the whole of classical mechanics. It is a specific, "Galilean"
theory of relativity - which then proved to be only approximate
and had to be replaced by another one, the ``Einsteinian'' theory.\\
\mbox{}\hfill{\it -- Jean-Marc L\'evy-Leblond}~\cite{Leblond1976}
\equo

From the perspective of pre-relativistic mechanics, the
freedom to use other functions for $T(p)$
than $T=p^2/(2\,m)$ seems to be of no use.
In the times of Galileo and Kepler, experiments would
have quickly lead to the conviction that terms beyond
second order in $T(p)$ have no physical relevance.
Today we know that this is wrong: The fact that $T(p)$
can - unlike $T(v)$ - not be completely determined from
the condition of energy conservation is indeed physically
significant. It allows to reconcile classical mechanics
with relativity.

There are various ways to derive the Lorentz transformations
and/or the relativistic energy-momentum-relation (REMR).
Many such derivations have been published, first by Einstein
arguing with the emission of light~\cite{Einstein1905},
later by many other authors ``without light'', for instance by
Terletskii~\cite{Terletskii1968}, L\"owdin~\cite{Loewdin98} and
L\'evy-Leblond~\cite{LevyLeblond1976}, to name just a few.
One of the shortest derivations of the REMR was given by
Davidon~\cite{Davidon1975}~\footnote{We became aware of
it by the papers of L\'evy-Leblond~\cite{Leblond1976,Leblond2003,Leblond2025}.}
and we shall briefly describe it here.

Once the concept of momentum is considered, one may define the
inertia $N$ of some physical body as the ratio of momentum to velocity
\begeq
N=p/v\,.
\label{eq_inertia}  
\endeq
In Newton's 'classical' case, this inertia is simply constant $N=m$.

However the experimental results of Kaufmann's cathode ray experiments
indicated an increase of the inertia with kinetic energy.
This experimental observation has a simple and direct formulation,
namely that the increase in the inertia $N$ of a body is proportional
to the increase of it's kinetic energy:
\begeq
dN=\chi\,dT
\label{eq_inertia2}
\endeq
with some (universal) constant proportionality factor $\chi$.
When combined with (Hamilton's) definition of velocity $dT=v\,dp$,
this yields
\myarray{
  dN&=\chi\,v\,dp=\chi\,v\,(N\,dv+v\,dN)\\
  {dN\over N}&={\chi\,v\over 1-\chi\,v^2}\,dv\\
  N&={N_0\over\sqrt{1-\chi\,v^2}}\,.
}
In the low-velocity limit $v\to 0$, the inertia must equal
the mass so that $N_0=m$ so that in this case $p=m\,v$ holds.
The constant $\chi$ has the dimension of an inverse squared
velocity and can be determined by experiment. This velocity
apparently corresponds to the velocity of light $c$ such that
$\chi=c^{-2}$ holds. Then one eventually obtains
\begeq
p={m\,v\over\sqrt{1-v^2/c^2}}
\endeq
Note that this derivation works only with the Hamiltonian
definition of velocity (Eq.~\ref{eq_vdef}).
With some algebra one then derives 
\begeq
v={p\,c^2\over\sqrt{m^2\,c^4+p^2\,c^2}}={dT\over dp}\,,
\endeq
and by integration one obtains the kinetic energy:
\begeq
T(p)=\sqrt{m^2\,c^4+p^2\,c^2}+C\,.
\endeq
Since $T(p)$ vanishes by definition for $p\to 0$, one
has to demand that $C=-m\,c^2$: 
\begeq
T(p)=\sqrt{m^2\,c^4+p^2\,c^2}-m\,c^2\,.
\label{eq_remr}
\endeq
The assumption (Eq.~\ref{eq_inertia2}), combined with
Hamiltonian mechanics as derived from the
principle of energy conservation, leads to the correct
relativistic kinematics. In this way one can derive
core equations of Einstein's special relativity on a single
page without the use of the light postulate and without
even mentioning the tedious and counter-intuitive
transformations between 'inertial frames'.
Furthermore one has a clear and unambigues
definition of the meaning of the notion of {\it inertia}
in mechanics: it is the ratio of momentum and velocity.

We do not suggest to abandon the discussion of transformations
between inertial frames. But one should always be aware of
the fact that these transformations are and will remain
``thought experiments'' as we have no practical means to
perform experiments with observers at ``relativistic''
velocities. Particles can be accelerated to relativistic
velocities with moderate technical effort. But no experiment
to date accelerated macroscopic objects to velocities which
deserve to be called relativistic. This does not mean that
we have doubts about the correctness of the Lorentz transformations
for macroscopic objects. But it is an oddity that a physical
theory of fundamental importance as Einstein's special
relativity is introduced almost entirely on the basis of
virtual experiments that are practically impossible to ever
be performed with real ``observers''.

There is yet another theoretical possibility to ``derive'' the
Dirac-Clifford algebra and in consequence the {\it form} of
Eq.~\ref{eq_remr} and the Lorentz transformations from an
algebraic analysis of a pure and abstract Hamiltonian
Ansatz~\cite{qed_paper,osc_paper}, i.e. from a ``pure''
universal conservation law.
However, a detailed discussion of this approach would exceed
the scope of this paper.

\subsection{Resolution of the Apparent Paradox}
\label{sec_paradox}

The two approaches given in Sec.~\ref{sec_xv} and Sec.~\ref{sec_xp} seem
to be in a logical conflict. Because, if the kinetic energy is a function
of momentum and the momentum is a function of the velocity only, then
the kinetic energy is still a function of velocity only. But then it
remains unclear how the first method can possibly yield a 'wrong', namely
Newtonian, result.

The solution of the apparent paradox lies in the (false) presumption that
the {\it acceleration} is independent of the velocity. This was presumed
in the Newtonian case, when we made use of Eq.~\ref{eq_1}. In the second
case however, the acceleration isn't explicitely used, but is replaced
by the change of momentum, which indeed {\it is} independent of velocity.

To see this more clearly, we compute the acceleration and
obtain, in the first case,
\begeq
a=-{1\over m}\,{dV\over dx}\,,
\endeq
which is a function of position only. In the second (``relativistic'')
case however one obtains
\begeq
a=-{(1-v^2/c^2)^{3/2}\over m}\,{dV\over dx}\,,
\endeq
which is a function of position {\it and} velocity. Hence there
is more to say about the change of the form of Newton's second law from
$F=m\,a$ to $F={dp\over dt}$ than is usually explained in standard
textbooks. The difference is marginal as long as $v\ll c$, but if the
velocity is close to $c$, as in many particle accelerators, then  
it becomes virtually impossible to {\it accelerate} particles, while
there is nothing that prevents us from raising the particle's
momentum further~\footnote{Hence it is, at very high energies, a misnomer to
speak of particle {\it accelerators}. It would be more appropriate
to call these machines particle {\it energizers}.}.

The replacement of velocity by momentum as (one of) the fundamental dynamical
variable(s) is the most obvious difference between Lagrangian and
Hamiltonian mechanics. As we tried to explain, this change of variable(s)
is the essential ingredient of ``relativistic'' mechanics and supports
the view that the theory of special relativity has it's simplest and
most natural formulation in the Hamiltonian rather than in the
Lagrangian formalism.

\section{Lagrangian Mechanics}
\label{sec_LM}

Lagrangian mechanics can now be introduced, if desired,
via the Legendre transform. This has been suggested
before~\cite{Hanc2004}, but without the (as we think)
required emphasis on the use of the momentum.
This is a reversal of the usual textbook order insofar as we suggest to derive
Lagrangian from Hamiltonian dynamics and not vice versa - and it provides some
evidence that the so-called ``principle of least action'' (PLA) is not required
to obtain the Lagrangian equation(s) of motion: Then the PLA is not an
input but an output of the theory.

One defines the Lagrangian function ${\cal L}(x,v)$ as follows:
\begeq
{\cal L}(x,v)=v\,p-{\cal H}(x,p)\,.
\endeq
where ${\cal H}(x,p)$ is the Hamiltonian function, which is
usually identical with the energy.
The momentum $p$ can then be obtained by
\begeq
p={\d{\cal L}\over\d v}\,.
\label{eq_pdef}
\endeq
The second partial derivative is then
\begeq
{\d{\cal L}\over\d x}=-{\d{\cal H}\over\d x}=-{d V\over\d x}
\endeq
and the equation of motion is therefore:
\begeq
{d p\over dt}={d\over dt}\left({\d{\cal L}\over\d v}\right)={\d{\cal L}\over\d x}\,,
\endeq
or, respectively:
\begeq
{d\over dt}\left({\d{\cal L}\over\d v}\right)-{\d{\cal L}\over\d x}=0
\label{eq_variation}
\endeq
As well known, it is this {\it last} equation that can be derived 
from the PLA. But Eq.~\ref{eq_variation} is of little use if the
Lagrangian function is unknown: The PLA does not provide us with
equations of motion unless we have means to guess ${\cal L}(x,v)$.
This is very different in case of energy conservation: Firstly the
principle and the form $E=T+V$ of energy conservation can be obtained
from the observation of simple physical systems like a pendulum.
Secondly, as demonstrated in Sec.~\ref{sec_xv}, the form of the
kinetic energy can be derived from $E=T+V$ and ${dE\over dt}=0$.
Even though this leads to classical Newtonian mechanics, it is
still rather  straightforward to make the transition to relativistic
mechanics. Neither $E=T+V$ nor ${dE\over dt}=0$ have to be replaced
for this transition. And the Legendre transform allows to derive
Langrangian mechanics from there in both, Galilean and Einsteinian,
relativity. The reverse is however not true.

Nonetheless it is tradition to introduce classical mechanics in
the confusing historical order; it invariably starts with Newton's
``laws'', proceeds with Lagrangian mechanics and finally arrives
via the Legendre transform at the Hamiltonian ``formulation''.
These three ``formulations'' of classical mechanics are often said
to be equivalent~\cite{Davis1986,Kelley2017} - which is highly
questionable as demonstrated above.

The Lagrangian is usually introduced as ${\cal L}=T-V$.
Attempts to motivate this choice of the Lagrangian have been made~\cite{Rojo2005},
despite the well-known fact that it is wrong in general, specifically
in (relativistic) mechanics proper. This classical definition ${\cal L}=T-V$
of the Lagrangian requires to move the goal posts when one wishes
to ``derive'' the relativistic equations of motion from a
Lagrangian. This is due to the Lagrangian definition of the momentum
(Eq.~\ref{eq_pdef}), which fails in the ``relativistic'' case for
${\cal L}=T-V$.

Despite all appraisal of the PLA: in practice no one ever derives
equations of motion from a Lagrangian. It is just the other way around:
Lagrangians are postulated when they are known to yield the correct
equations of motion.
It should be acknowledged, however, that the goal posts remain unchanged
when classical mechanics is taught not in historical but in logical
order, namely if the Lagrangian is derived from the Hamiltonian as
suggested here:
\myarray{
  {\cal L}(v)&=v\,p-m\,c^2\,(\y-1)\\
  &=m\,\y\,v^2-m\,c^2\,(\y-1)\\
  &=m\,c^2-m\,c^2\,{1-v^2/c^2\over\sqrt{1-v^2/c^2}}\\
  &=m\,c^2-m\,c^2\,\sqrt{1-v^2/c^2}\\
}
The constant term may be skipped to give the correct relativistic
Lagrangian of a free particle:
\begeq
{\cal L}(v)=-m\,c^2\,\sqrt{1-v^2/c^2}\,.
\endeq
Lagrangians have their well-deserved place in physics. But the
PLA and other ``variational principles'' should not any more
be regarded as foundational.

Most textbook authors, also distinguished and renowned
physicists like Feynman and co-authors~\cite{FLS}, introduce
Lagrangian mechanics from the PLA, which is often depicted as
an ``absolutely fascinating'' and ``deep'' principle of nature.
Such claims echo the theological and/or philosophical
motivations of some scientists of the 18th and 19th centuries, who
desired to find a proof that our world is the best of all
possible worlds. This latter motivation and attitude is widely
disregarded as untenable today~\cite{Yourgrau}:
\bquo
Obsessed with the urge to achieve "cognition
of real nature," i.e., to gain an "explanation" of material phenomena,
physicists have assigned to a particle the knowledge which enables
it to take the most convenient path of all those which would direct it to
its destination. The Aristotelian final cause is dragged into the subject,
and the particle is presented to us as what Poincare calls "a living and
free entity." Aristotle's dictum that nature does nothing in vain may have
been at the foot of Maupertuis' teleological principle, which initiated
the study of variational principles in nature, but to the twentieth-century
physicist the notion of such a mysterious purposive agency would only be a
regress into sheer obscurantism. History knows of many instances where a
science had its beginnings in a mythical and magical symbolism which
gradually shed its mystical features as our rational knowledge increased
and our conceptions became clearer- witness the transition from astrology to
astronomy and from alchemy to chemistry.
\equo
And despite many claims to the contrary it has never been proven that the
set of all ``possible worlds'' has more than a single element.

In contrast to the questionable claim that the PLA constitutes something
like an ``optimal'' world, energy conservation is not a property of an
``optimal'' but rather of a {\it real} world. It's trivialized message
is that there is no free lunch: If you want work to be done, you have to
pay with energy for it~\cite{French1986}:
\bquo
Energy is what we have to pay for in order to get things done.
[...] We do not think in terms of paying for force, or acceleration, or momentum.
Energy is the universal currency that exists in apparently countless denominations;
and physical processes represent a conversion from one denomination to another.
\equo
In the view of the author it is a necessary truth in {\it any} real world.

\section{Conclusions}
\label{sec_disc}

It has been criticized that~\cite{Grayson2006}
\bquo
[...] physics curricula typically consist of topics whose selection
and ordering are guided more by habit or tradition than by
cognitive research or sound pedagogy.
Fundamental principles and the hierarchical nature of
physics knowledge are often not highlighted, with the result
that students perceive physics as an impossibly large
body of facts and equations to be memorized.
\equo
As we have shown, a systematic approach, based on simple principles
that can be derived from observation and experiment, is available.
Energy conservation can easily be illustrated using simple mechanical
systems, like for instance a pendulum or a spring-mass-system,
while Newtonian mechanics (as it is usually taught) requires an
abstract and logically questionable notions of force-free bodies
moving in absolute space~\cite{Rigden1987}:
\bquo
The first law [...] is a logician’s nightmare. [...] To teach Newton’s laws so that
we prompt no questions of substance is to be unfaithful to the discipline itself.
\equo
If Rigden is right, then the prevailing physics curriculum, notorious
for the Newton-first-approach, repels students skilled in logic - literally by design.

The usual textbook introduction to mechanics seems to be but an excercise in
epigonism: Almost invariably common textbook mechanics begins with Newton's ``laws'',
``derives'' Langrangian mechanics from it and finally derives Hamiltonian
from Lagrangian mechanics. This is the standard approach and only a small
minority of textbooks and lecture notes deviates from this traditional order.
It is rarely emphasized that Lagrangian is more general than Newtonian
mechanics and Hamiltonian is more general than Lagrangian mechanics.
A logical approach starts with the most general and fundamental principles
and then derives the more specific from the more general laws.
Unless one wishes to teach history (of physics) and not physics proper,
the order should be reversed and energy conservation should be used
to derive classical mechanics.

Eventually Hertz' 125 year old outlook might (some day) become good
practice in physics teaching~\cite{Hertz1899}:
\bquo
Now, towards the end of the century, physics has shown a 
preference for a different mode of thought. Influenced by the 
overpowering impression made by the discovery of the principle
of the conservation of energy, it likes to treat the 
phenomena which occur in its domain as transformations of 
energy into new forms, and to regard as its ultimate aim the 
tracing back of the phenomena to the laws of the transformation
of energy. This mode of treatment can also be applied 
from the beginning to the elementary phenomena of motion. 
There thus arises a new and different representation of 
mechanics, in which from the start the idea of force retires in 
favour of the idea of energy.
\equo

\bibliographystyle{unsrturl}
\bibliography{wnm_paper.bib}

\end{document}